\documentclass[twocolumn,preprintnumbers,amsmath,amssymb,nofootinbib]{revtex4}

\usepackage{epsfig}
\usepackage{epstopdf}
\usepackage{graphicx}
\usepackage{dcolumn}
\usepackage{bm}
\usepackage{subfigure}
\usepackage{booktabs}

\begin{document}

\title{Testing Isotropy in the Local Universe}

\author{${}^{a}$Stephen Appleby, ${}^{a,b}$Arman Shafieloo}
\affiliation{${}^{a}$Asia Pacific Center for Theoretical Physics, Pohang, Gyeongbuk 790-784, Korea \\
${}^{b}$Department of Physics, POSTECH, Pohang, Gyeongbuk 790-784, Korea}

\date{\today}

\begin{abstract}
We test the isotropy of the local distribution of galaxies using the 2MASS extended source catalogue. By decomposing the full sky survey into distinct patches and using a combination of photometric and spectroscopic redshift data, we use both parametric and non-parametric methods to obtain the shape of the luminosity function in each patch. We use the shape of the luminosity function to test the statistical isotropy of the underlying galaxy distribution. The parametric estimator shows some evidence of a hemispherical asymmetry in the north/south Galactic plane. However the non-parametric estimator exhibits no significant anisotropy, with the galaxy distribution being consistent with the assumption of isotropy in all regions considered. The parametric asymmetry is attributed to the relatively poor fit of the functional form to the underlying data. When using the non-parametric estimator, we do find a dipole in the shape of the luminosity function, with maximal deviation from isotropy at galactic coordinate $(b,l)=(30^{\circ},315^{\circ})$. However we can ascribe no strong statistical significance to this observation. 
\end{abstract}

\maketitle % --------------------------------------------------------------------

\section{Introduction} 

The large scale homogeneity and isotropy of spacetime are two fundamental tenets underpinning modern cosmology. The assertion that spacetime, when averaged over sufficiently large length scales, can be described in terms of an FLRW metric and a system of perfect fluids has proved to be extremely successful; the standard $\Lambda$CDM model fits all current cosmological data sets satisfactorily (although certain curiosities remain, see \cite{Ade:2013lmv,Battye:2013xqa,Verde:2013wza,Efstathiou:2013via,Colin:2010ds,Feindt:2013pma,Appleby:2013ida}). Isotropy on large scales is strongly supported by the Cosmic Microwave Background (CMB), specifically in the uniformity of the distance to last scattering on the sky \cite{Campanelli:2006vb,Campanelli:2007qn,Appleby:2009za,Campanelli:2011uc}. Similarly current galaxy surveys have been shown to approach a homogeneous distribution when averaged over sufficiently large volumes \cite{Hogg:2004vw,Sarkar:2009iga,Scrimgeour:2012wt,Alonso:2013boa}. 

Despite the formidable success of the standard model, there remain unresolved issues. Perhaps the most interesting are the CMB anomalies detected on large scales \cite{Hinshaw:1996ut,Spergel:2003cb,Copi:2010na,Ade:2013nlj,Akrami:2014eta}, in particular a deficiency of large angle correlations. It is not simply the lack of correlation on large scales that is of interest. Measurements of the CMB quadrupole and octopoles show an alignment approximately perpendicular to the ecliptic plane \cite{de OliveiraCosta:2003pu,Abramo:2006gw,Land:2005ad,Land:2006bn,Rakic:2007ve,Samal:2007nw,Samal:2008nv,Eriksen:2007pc,Hoftuft:2009rq,Copi:2006tu,Copi:2005ff,Schwarz:2004gk,Souradeep:2006dz}
. Such violations of isotropy persist to smaller scales $l \leq 60$. The statistical significance of these observations has been widely debated in the literature, since most constitute an {\it a posteriori} search for anomalies in the data, which in turn is thought to be a single realisation of a Gaussian and isotropic distribution. 

The large scale nature of these deviations from isotropy could suggest a cosmological origin. If this were the case, then one might expect to observe a similar signal in other cosmological data sets \cite{Fernandez-Cobos:2013fda,Cai:2013lja}. In this work we attempt to quantify the extent to which current Large Scale Structure (LSS) data sets are consistent with the underlying assumptions of isotropy and homogeneity. Curiosities have already been observed in the local galaxy distribution, not least recent work \cite{Keenan:2009jh,Keenan:2012gr,Keenan:2013mfa,Whitbourn:2013mwa} (see also related earlier works \cite{Frith:2003tb,Busswell:2003ta,Frith:2005az,Frith:2004wd,Frith:2004tw}) which indicates that the local luminosity density is systematically lower than expected, out to co-moving distances $d_{\rm c} \sim 300 {\rm Mpc}$.

In this work, we analyse the local galaxy distribution using the 2MASS extended source catalogue \cite{Skrutskie:2006wh}, using the photometric redshift data developed recently in \cite{Bilicki:2013sza}. Spectroscopic redshifts are provided in the 2MASS catalogue for $\sim 45,000$ galaxies to limiting magnitude $K_{\rm s} < 11.75$, where $K_{\rm s}$ is the K-band magnitude; this sample has a mean redshift of approximately $\bar{z} \sim 0.028$. Using photometric redshift data for the full extended source galaxy catalog allows us to probe a considerably deeper volume - the mean redshift of a sample cut at $K_{\rm s} < (13.5, 13.9)$ is approximately $\bar{z} \sim (0.075,0.1)$ respectively. However the gain in volume is offset by the increased uncertainties associated with photometric redshift data.

Using a full sky galaxy survey allows us to examine the angular distribution of galaxies on the sky, and the volume probed in the full 2MASS catalogue allows us to make magnitude cuts to test the redshift distribution. In what follows we build a picture of the three dimensional galaxy distribution using the luminosity function. We are specifically interested in the existence of preferred directions in the data, which are indicative of a breakdown of statistical isotropy. We analyse the galaxy distribution in disjoint patches in the sky, using a combination of parametric and non-parametric methods to reconstruct the luminosity function. The shape of the luminosity function in different regions of the sky provides us with a test of the isotropy of the local Universe. 

The paper will proceed as follows. We outline the method we use to construct the luminosity function in section \ref{sec:2}. In \ref{sec:3} we describe the data set used and discuss our choice of quality cuts. Our results are presented in section \ref{sec:4}, along with a discussion of our error analysis. We highlight some of the limitations of our method  and conclude in section \ref{sec:5}. Unless otherwise stated, throughout this work we use cosmological parameters $\Omega_{\rm m0}=0.27$, $\Omega_{\rm k} = 0$ and take $H_{\rm 0} = 100 h \: {\rm km} \: {\rm s}^{-1} \: {\rm Mpc}^{-1}$.

\section{\label{sec:2}Luminosity Function}

The aim of this work is to determine the angular dependence of the local galaxy distribution. To do so, we calculate the luminosity function $\phi(L)$, the number of galaxies per unit luminosity $L$ per unit volume \cite{Johnston:2011bx,Takeuchi:2000xh}. To obtain the luminosity function, one can use either parametric or non-parametric methods. The overwhelmingly favoured parametric form used in the literature is the Schechter function \cite{Schechter:1976iz}

\begin{equation}\label{eq:2} \phi(L) dL = {\phi^{*} \over L^{*}} \left({L \over L^{*}}\right)^{\alpha} \exp\left[-{L \over L^{*}}\right] {dL \over L^{*}} \end{equation} 

\noindent which contains three parameters - the density normalisation ($\phi^{*}$), the power law slope at low $L$ ($\alpha$) and the typical luminosity at the knee of the function ($L^{*}$). The expression ($\ref{eq:2}$) is more commonly written in terms of absolute magnitude $M$ via the relation

\begin{equation} \label{eq:3} M - M^{*}  = -2.5 \log_{10}\left({L \over L^{*}}\right) \end{equation}

\noindent yielding 

\begin{widetext}
\begin{equation}\label{eq:4} \phi(M) dM = 0.4\ln[10] \phi^{*} 10^{0.4(1+\alpha)(M-M^{*})} \exp\left[ -10^{0.4(M-M^{*})}\right] dM \end{equation}
\end{widetext}

\noindent For the survey that we use, each galaxy is characterized by a photometric (or spectroscopic for a small subset) redshift and three apparent magnitude measurements, made in the $(H,J,K_{\rm s})$ bands. We utilize the $K_{\rm s}$ band in what follows, specifically taking the $20 {\rm mag}$ ${\rm arcsec}^{-2}$ isophotal fiducial elliptical aperture magnitudes. For a galaxy with redshift and apparent magnitude $(z_{\rm i},m_{\rm i})$, we convert from apparent to absolute magnitude $M_{\rm i}$ using 

\begin{equation}\label{eq:mag} M_{\rm i} = m_{\rm i} - 5\log_{\rm 10} \left[ d_{\rm l}(z_{\rm i})\right]-25 + K(z_{\rm i},K_{\rm s,i},J_{\rm i}) + E(z) \end{equation}

\noindent where $K(z_{\rm i},K_{\rm s,i},J_{\rm i})$ is the K-correction for the galaxy - which accounts for the effect of redshift on the bandpass transmission curve. We follow \cite{Chilingarian:2010sy} to construct this function, using the following approximate fitting form 

\begin{equation} \label{eq:Kc} K(z,K_{\rm s},J) = \sum_{k,j} a_{\rm k,j} z^{k} (J-K_{\rm s})^{j} \end{equation}

\noindent where the $a_{\rm k,j}$ coefficients are given in \cite{Chilingarian:2010sy}. In \cite{calc} an online calculator was created to accurately calculate the K-corrections for a number of surveys - we use the 2MASS $K_{\rm s}$ correction with color $J2-Ks2$ in what follows. We repeated our analysis using a simple ansatz $K(z) = -6\log [1+z]$ for the K-correction \cite{Kochanek:2000im}, and found no significant effect on our results. As discussed in \cite{Mannucci:2001qa}, the form of the K-correction at low redshifts is nearly independent of the galaxy type. Furthermore, it has been shown that the K-correction specified by ($\ref{eq:Kc}$) is consistent with more rigorous calculations to within $\sim \Delta M = 0.1{\rm Mag}$ \cite{Fioc:1997sr,Blanton:2006kt}. Following \cite{Blanton:2002wv} we adopt a simple ansatz $E(z) = Qz$ with $Q=1$ to describe the redshift evolution.

To fit the model parameters $(\phi^{*},M^{*},\alpha)$, one must use a maximum likelihood estimation technique (the so called STY method \cite{Sandage:1979re}). If we have a sample of $N_{\rm gal}$ galaxies in a magnitude limited survey, then the probability that the $i^{\rm th}$ galaxy has luminosity $M_{\rm i}$ given it is located at redshift $z_{\rm i}$ is given by \cite{Sandage:1979re}

\begin{equation} \label{eq:5} p_{\rm i} = P(M_{\rm i}|z_{\rm i}) = {\phi(M_{\rm i}) \over \int_{M_{\rm min}(z_{\rm i})}^{M_{\rm max}(z_{\rm i})} \phi(M') dM' } \end{equation}

\noindent where $M_{\rm max,min}$ are the maximum and minimum absolute magnitudes at redshift $z_{\rm i}$ that could be detected in the magnitude limited sample. The likelihood 

\begin{equation}\label{eq:6} {\cal L} = \Pi_{\rm i=1}^{N_{\rm gal}} p_{\rm i} \end{equation}

\noindent is then maximized for the parameters $(M^{*},\alpha)$. As each individual probability $p_{\rm i}$ is given by a ratio of the luminosity function and its integral over an absolute magnitude range, the normalization $\phi^{*}$ drops out of the calculation when using this method.  

The Schechter function has been shown to be a relatively poor fit the observed galaxy distribution - it has been argued \cite{Jones:2006xy} that it falls off too sharply at the bright end and hence fails to simultaneously fit both the bright and faint tails. Care must therefore be taken that any anisotropic signal that we detect is not spurious, due to the functional form chosen yielding a better or worse fit in different patches of the sky. 

With this in mind, we also attempt to reconstruct the luminosity function using a non-parametric technique. Numerous methods have been developed for this purpose - we direct the interested reader to \cite{Takeuchi:2000xh} for a review. Here we adopt the stepwise maximum likelihood method introduced by Efstathiou-Ellis-Peterson (EEP) \cite{Efstathiou:1988pa}.

The EEP method estimates the binned luminosity function 

\begin{equation} \phi(M) = \sum_{i=1}^{N_{\rm bin}} \phi_{\rm i} W(M_{\rm i} - M) \end{equation} 

\noindent with window function 

\begin{equation} W(M_{\rm i} - M) \equiv 
\begin{cases}
    1,& \text{if } M_{\rm i} - \Delta M/2 \leq M \leq M_{\rm i} + \Delta M/2 \\
    0,              & \text{otherwise}
\end{cases}
\end{equation} 

\noindent and we take $N_{\rm bin}$ magnitude bins. The likelihood function is given by 

\begin{equation}\label{eq:b1} {\cal L} = \Pi_{i=1}^{N_{\rm gal}} {\sum_{j=1}^{N_{\rm bin}} W(M_{\rm j}-M) \phi_{\rm j} \over \sum_{k=1}^{N_{\rm b}} \phi_{\rm k} H(M_{\rm lim}(z_{\rm i}) - M_{\rm k}) \Delta M } \end{equation}

\noindent where $\Delta M$ is the width of the magnitude bins (here we adopt constant spacing), $M_{\rm lim}(z)$ is the maximum absolute magnitude that can be detected at redshift $z$ given the magnitude limits of the survey, and 

\begin{widetext}
\begin{equation} H(M_{\rm lim}(z_{\rm i}) - M) \equiv 
\begin{cases}
    1,& \text{if } M_{\rm lim}(z_{\rm i})-\Delta M/2 > M \\
    {M_{\rm lim}(z_{\rm i})-M \over \Delta M} + {1\over 2},     & \text{if } M_{\rm lim}(z_{\rm i})-\Delta M/2 \leq M < M_{\rm lim}(z_{\rm i})+\Delta M/2 \\
    0,              & \text{if } M_{\rm lim}(z_{\rm i})+\Delta M/2 \leq M  
\end{cases}
 \end{equation} 
\end{widetext}

\noindent To obtain the luminosity function, we minimize ($\ref{eq:b1}$) with respect to $\phi_{\rm k}$. It can be shown that the problem reduces to solving the equation 

\begin{equation} \phi_{\rm k} \Delta M = {\sum_{i=1}^{N_{\rm gal}} W(M_{\rm k} - M_{\rm i}) \over \sum_{\rm j=1}^{N_{\rm gal}} { H(M_{\rm lim}(z_{\rm j})-M_{\rm k}) \over \sum_{l=1}^{N_{\rm bin}} \phi_{\rm l} H(M_{\rm lim}(z_{\rm j})-M_{\rm l})\Delta M} } , \end{equation}

\noindent which can be solved by iteration. As an initial guess, we use the Schechter function with best fit parameters estimated using the STY method. We have tested that the choice of initial guess has no effect on the shape of the final binned luminosity function. 

Both the non-parametric EEP and parametric STY methods yield no information regarding the overall density of the galaxy distribution, as both involve ratios of the luminosity function resulting in estimates that are invariant under amplitude shifts. However, in this work we are purely interested in the the shape of the luminosity function, which we use as a test of the statistical isotropy of the galaxy distribution. The normalisation $\phi^{\ast}$ will give us information regarding the homogeneity of the low redshift data - local fluctuations will have the effect of shifting the amplitude of $\phi(M,z)$ in different patches with respect to one another. Estimates of the amplitude typically depend on the volume of the sample, and hence are strongly sensitive to photometric redshift uncertainties. Large errors in the normalisation $\phi^{\ast}$ would lessen the significance of any anisotropic signal detected. In a companion paper, we explore the homogeneity of the local galaxy distribution by using a series of non-parametric estimators to calculate the galaxy density as a function of magnitude cut. For the purposes of this work, we focus on the shape of the luminosity function.

\section{\label{sec:3}Data - 2MASS Extended Source Catalog}

In this work we exclusively use the 2MASS all sky extended source catalogue (XSC) \cite{Skrutskie:2006wh}. 2MASS is an all sky, near-infrared survey that scanned more than $99.99\%$ of the sky during its four year operation. The resulting catalog contains more than 400 million point and 1.6 million extended sources - in the latter case the majority are extra-galactic. The measurements were performed over three bands - $J, (1.25\mu{\rm m})$ , $H, (1.65\mu{\rm m})$ and $K_{\rm s}, (2.16\mu{\rm m})$, with the signal to noise limit $S/N > 10$ being met by objects brighter (or of equal brightness) than $K_{\rm s} = 13.5{\rm mag}$ in the $K_{\rm s}$ band. The reliability of the survey is greater than $99\%$ in regions $|b|>20^{\circ}$, where $b$ is the galactic latitude. The photometry in the north and south galactic plane is highly symmetric, with a mean color difference of less than $\sim 0.01 {\rm mag}$. 

In what follows we use the newly constructed 2MASS XSC photometric redshift catalog - 2MPZ \footnote{http://surveys.roe.ac.uk/ssa/TWOMPZ} - constructed in \cite{Bilicki:2013sza}. We sketch the catalog here, interested readers should consult \cite{Bilicki:2013sza} for further details. The full XSC catalog contains $1,646,966$ objects, however before use we must remove a small number due to measurement errors and the existence of artifacts. After removing artifacts, non-extended Galactic sources, extreme $K_{\rm s},J,H$ measurements and those objects lacking measurements in the $(J,H)$ bands, one arrives at the `2MASS good' sample with $1,471,442$ entries. The $K_{\rm s},J,H$ $20 {\rm mag}$ ${\rm arcsec}^{-2}$ isophotal magnitudes were corrected for Galactic extinction with the standard SFD maps \cite{Schlegel:1997yv}, with coefficients $A_{\lambda}/E(B-V)$ taken from \cite{Cardelli:1989fp}. Spectroscopic redshifts were given in cases where known, either from the 2MASS redshift catalog with spectroscopic redshift data provided to limiting magnitude $K_{\rm s} < 11.75$, or from other publicly available data sets (specifically Sloan Digital Sky Survey Data Release 9, 6DF Galaxy Survey Data Release 3, 2DF Galaxy Redshift Survey and the ZCAT compilation using the CFA Redshift catalog \cite{Colless:2001gk,Huchra:2011ii,Jones:2004zy,Ahn:2012fh,Eisenstein:2011sa,Desai:2012zr,Wright:2010qw,Huchra:1996}). 

Approximately one third of the galaxies in the 2MPZ catalog have spectroscopic redshifts. The remaining galaxies are assigned photometric redshifts using a machine learning technique \cite{Firth:2002yz}, which is based upon an empirical relation between galaxy magnitudes and redshifts obtained using the spectroscopic redshift subsample as a training set. Other methods, such as template (Spectral Energy Distribution (SED)) fitting \cite{Bolzonella:2000js}, are also powerful tools to infer photometric redshift information. However, such methods are not appropriate for the 2MPZ catalog as the SuperCOSMOS magnitudes are obtained from photographic plates and are not easily calibrated to the same system as 2MASS and WISE \footnote{We would like to thank Maciej Bilicki for clarification on this point}. The potentially problematic photometric calibrations only apply to the SuperCOSMOS optical data, and could be translated into a small offset in photometric redshifts (although the effect is expected to be negligible). No such issues occur for the 2MASS K-band photometry that we use in this work. The 2MPZ catalog was constructed using the publicly available ANNz package - for a description of the algorithm adopted we direct interested readers to \cite{Collister:2003cz}.

\begin{figure}
\centering
\mbox{\resizebox{0.4\textwidth}{!}{\includegraphics[angle=270]{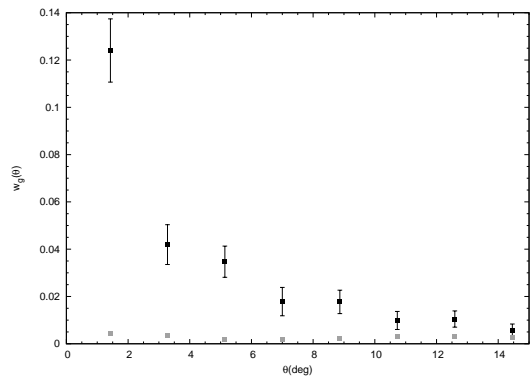}}}
\caption{The galaxy auto-correlation function is exhibited as a function of angular bin (in degrees) (black squares), along with error bars obtained via jackknife re sampling. The cross correlation function of the galaxy distribution and the 2MASS point source catalog is shown as grey dots. We have applied the same mask $|b| < 20^{\circ}$ to both galaxy and point source maps. The cross correlation is clearly sub-dominant for our choice of galactic cut, indicating that our galaxy sample is free from stellar contamination.  
}
\label{fig:1} 
\end{figure}

When testing the isotropy and homogeneity of the galactic distribution, one typically requires a complete and uniform sample over the survey volume under consideration - for this reason one must apply a magnitude cut to the 2MASS data. We adopt a slightly conservative cut of $K_{\rm s} < 13.5$ band to ensure completeness over $\sim 96\%$ of the sky (excluding the galactic bulge). $531,877$ galaxies remain in the sample with this cut. For the purposes of our analysis, we also cut the region $|b|<20^{\circ}$ from the sky. This is principally to avoid stellar contamination, which has been reported at lower galactic latitudes \cite{Maller:2003eg}. In fig.\ref{fig:1} we exhibit the cross correlation of the XSC with the 2MASS Point Source Catalog (PSC) for galactic cut $|b| < 20$ - we find the cross correlation to be completely sub-dominant compared to the auto-correlation (shown in black). The error bars are obtained via jack knife re-sampling.

\begin{figure}
\centering
\mbox{\resizebox{0.4\textwidth}{!}{\includegraphics[angle=270]{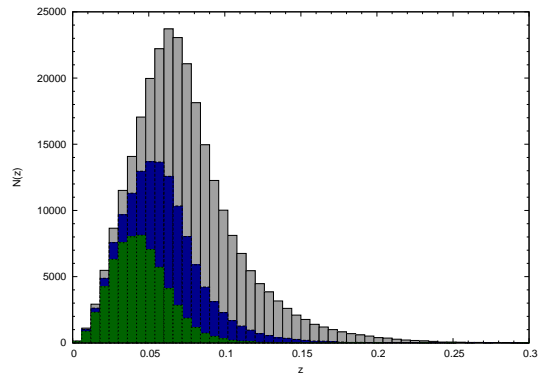}}}
\caption{The distribution of 2MASS galaxies as a function of redshift, for three magnitude cuts $K_{\rm s} < 12.5, 13.0, 13.5$. Successively large magnitude cuts allow us to probe an increasingly large volume, out to mean redshift $\bar{z} \sim 0.074$ in this work. 
}
\label{fig:2} 
\end{figure}

In fig.\ref{fig:2} we exhibit the galaxy distribution as a function of redshift for a number of $K_{\rm s}$ cuts. The mean redshift of the survey is given by $\bar{z}=(0.036,0.045,0.074)$ for $K_{\rm s} \leq (12.0,12.5,13.5)$ respectively.

\subsection{\label{sec:err}Error Estimation}

To assess the effect of photometric redshift uncertainties on the shape of the luminosity function, we construct $N_{\rm real} =1000$ realisations of the galaxy catalog and repeat our analysis. The mocks are constructed using the following algorithm. 

In each patch of the sky, we bin the galaxies into $N_{\rm bin}=20$ magnitude bins between $M = (-20,-26){\rm mag}$ - each bin contains $N_{\rm gal, i}$ galaxies, with $i=(1,20)$. For a single realisation, each magnitude bin is then assigned $\tilde{N}_{\rm gal,i}$ galaxies, where $\tilde{N}_{\rm gal, i}$ is drawn from a Poisson distribution of mean $N_{\rm gal, i}$. These galaxies are assigned redshifts and $K_{\rm s}$-band magnitudes $(z,K_{\rm s})$, drawn with replacement from the original $N_{\rm gal, i}$ galaxies in that bin. If the drawn galaxy has a quoted spectroscopic redshift $z_{\rm spec}$ in the catalog, then we simply assign the new data point this value - $(z_{\rm spec},K_{\rm s})$. If the drawn galaxy only has a photometric redshift associated with it, then we must infer its true (spectroscopic) value.

To estimate the spectroscopic redshift of a galaxy, we extract the subset of galaxies in the catalog that have both photometric $z_{\rm photo}$ and spectroscopic $z_{\rm spec}$ redshifts. We bin this sub-sample in $N_{\rm z} = 20$ {\it photometric} redshift bins, with limits chosen such that each bin has an equal number of galaxies. In each bin, we construct an empirical Probability Distribution Function (PDF) of $z_{\rm spec}-z_{\rm photo}$. Then the `true' spectroscopic redshift of a galaxy with only photometric redshift information is drawn from these PDF's. Once we have inferred $z_{\rm spec}$, we then calculate the data point's new absolute magnitude according to ($\ref{eq:mag}$).

\section{\label{sec:4}Results}

\begin{figure}
\centering
\mbox{\resizebox{0.45\textwidth}{!}{\includegraphics[angle=270]{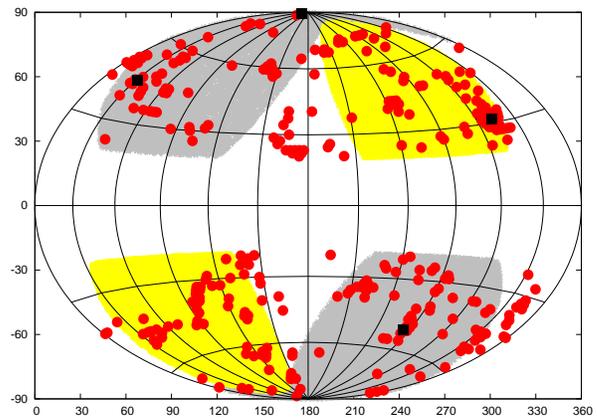}}}
\caption{After pixelizing the sphere into $N_{\rm pix} =12288$ equal area regions, we exhibit the pixels containing the $\sim 3.5\%$ largest galaxy number density as red dots. The large black squares indicate the most extreme peaks in the galaxy number counts, which correspond to the Shapley, Coma, Hercules and Horologium superclusters. We also exhibit patches C and D as grey shaded regions, and A and B in yellow. 
}
\label{fig:ait} 
\end{figure}

The primary purpose of this work is to locate and quantify any directional dependence in the local matter distribution. The theoretical tool that we use for this purpose is the luminosity function described in section \ref{sec:2}. To achieve our aim we decompose the sky into four distinct and orthogonal patches on the sky, centered at galactic coordinates $(b,l)=(38^{\circ},263^{\circ}), (-38^{\circ},83^{\circ}), (52^{\circ},83^{\circ}), (-52^{\circ},263^{\circ})$. We name these A,B,C,D respectively in what follows. Patch A and B have centers in the vicinity of the CMB dipole, and patches C and D were chosen such that they are perpendicular to patches A and B and maximally distant from our galactic latitude cut. The fractional area of patches A and B on the unit sphere is $\omega = 0.104$, and patches C and D have a larger value $\omega = 0.130$ by virtue of their distance from the galactic cut $|b| < 20^{\circ}$. Here $\omega = \Omega_{\rm patch}/4\pi$, where $\Omega_{\rm patch}$ is the area on the unit sphere that any given patch would cover. We note that the decision to make patches A,C and B,D different area on the sky was a conscious one by the authors -  it is our intention to show that the results obtained here are independent of the patch size and number of galaxies. We consider only galaxies with $z < 0.3$, and $8{\rm mag} < K_{\rm s} < 13.5 {\rm mag}$ - the number of galaxies in each patch with $K_{\rm s} < 13.5 {\rm mag}$ is given by $N_{\rm patch} = (3.1,2.9,3.6,3.7)\times 10^{4}$ for A-D respectively. We cut all galaxies with $z < 0.005$ from our sample. This removes objects for which the peculiar velocity is a dominant contributor to the redshift. It also removes extremely local features in the galaxy distribution. Patches A and C (B and D) are located in the north (south) galactic plane. In fig.\ref{fig:ait} we exhibit the four patches in galactic coordinates. We also highlight the regions of highest galactic density as red dots, and the positions of the largest superclusters in the sample as black squares (we also exhibit the approximate location of these superclusters in table \ref{tab:super}).

We calculate the luminosity function in these four patches, searching for statistically significant deviations in the shape of $\phi(M)$. Detailed studies of the local luminosity function have long established that the Universe is under-dense at distances $d_{\rm c} \sim 300 {\rm Mpc}$ \cite{Frith:2003tb}. In a follow up paper we assess the homogeneity of the local Universe using this galaxy sample, building upon recent studies of low redshift homogeneity \cite{Keenan:2012gr,Keenan:2013mfa}. However in this work we wish to quantify the statistical isotropy of the distribution.

%%%%%%%%%%%%%%%%%%%%%%% 
\begin{table}[!htb]
\begin{tabular}{l|ccc} 
Name \ & \  z  \ & b & \ l \  \\
\hline 
Shapley & $0.046$ & $+32^{\circ}$ & $+310^{\circ}$  \\ 
Coma & $0.023$ & $+87^{\circ}$ & $+58^{\circ}$  \\ 
Horologium & $0.067$ & $-54^{\circ}$ & $+262^{\circ}$  \\ 
Hercules & $0.037$ & $+45^{\circ}$ & $+30^{\circ}$  \\ 
\end{tabular}
\caption{The largest superclusters detected in the 2MASS sample, exhibited in fig.\ref{fig:ait} as black squares. The superclusters are extended objects and we quote their approximate positions only. The Abell and Ursa major superclusters are also present in the red sample of fig.\ref{fig:ait}.
}
\label{tab:super}
\end{table}

In each of the four patches, we have sufficiently many galaxies to also take magnitude cuts. We take successively $K_{\rm s} < 12.5,13.0,13.5 {\rm mag}$, keeping the lower limit as $K_{\rm s}  > 8 {\rm mag}$. We choose not to vary the lower magnitude limit since this would have the effect of removing large numbers of galaxies for which spectroscopic redshifts are known. By making successive magnitude cuts, we are probing a successively larger volume as indicated by fig.\ref{fig:2}.

\begin{widetext}
\begin{center}
\begin{table}
\begin{tabular}{c|ccccc|ccccc|ccccc}
\toprule 
    Patch & \multicolumn{5}{c}{$K_{\rm s} <12.5$} & \multicolumn{5}{c}{$K_{\rm s} < 13.0$} & \multicolumn{5}{c}{$K_{\rm s} < 13.5$} \\
    & $\alpha$ & $\delta \alpha$ & $M^{*}-5\log[h]$ & $\delta M^{\ast}$ & $\chi^{2}$ & $\alpha$ & $\delta \alpha$ & $M^{*}-5\log[h]$ & $\delta M^{\ast}$ & $\chi^{2}$ & $\alpha$ & $\delta \alpha$ & $M^{*}-5\log[h]$ & $\delta M^{\ast}$ & $\chi^{2}$  \\
    \midrule 
A     & -0.95 & 0.038   & -23.53  & 0.030  & 11.2   & -0.95  & 0.025  &-23.55  & 0.020 & 16.3  & -0.94  & 0.017  & -23.60  & 0.013  & 32.5   \\
B     & -0.94 & 0.043   & -23.41  & 0.032  & 15.0   & -0.90  & 0.028  &-23.44  & 0.021 & 18.6  & -0.90  & 0.020  & -23.52  & 0.015  & 32.8    \\
C     & -0.93 & 0.042   & -23.46  & 0.029  & 12.0   & -0.92  & 0.028  &-23.49  & 0.020 & 21.0  & -0.97  & 0.016  & -23.59  & 0.013  & 44.9   \\
D     & -1.01 & 0.041   & -23.53  & 0.030  & 10.3   & -0.92  & 0.026  &-23.50  & 0.018 & 23.6  & -0.84  & 0.018  & -23.51  & 0.013  & 50.0    \\
  \bottomrule
\end{tabular}
\caption{The best fit values of the Schechter function parameters $(\alpha,M^{\ast})$ for the four patches A-D, taking three magnitude cuts $K_{\rm s} < (12.5,13.0,13.5) {\rm mag}$. We observe some trends in the results - a consistently lower value of $M^{\ast}$ with increasing $K_{\rm s}$, and an apparent dichotomy in $M^{\ast}$ in the north-south galactic plane (patches A,C and B,D respectively). $\delta \alpha$ and $\delta M^{\ast}$ indicate the $1-\sigma$ errors on the parameters obtained from the STY maximum likelihood method outlined in section \ref{sec:2}. The $\chi^{2}$ values quoted in each patch and magnitude bin are obtained by comparing the Schechter function fit to the non-parametric estimation of $\phi$ in $N_{\rm bin} =20$ bins, and are discussed further in section \ref{sec:4}.}
\label{tab:1}
\end{table}
\end{center}
\end{widetext}

\begin{figure}
\centering
\mbox{\resizebox{0.4\textwidth}{!}{\includegraphics[angle=0]{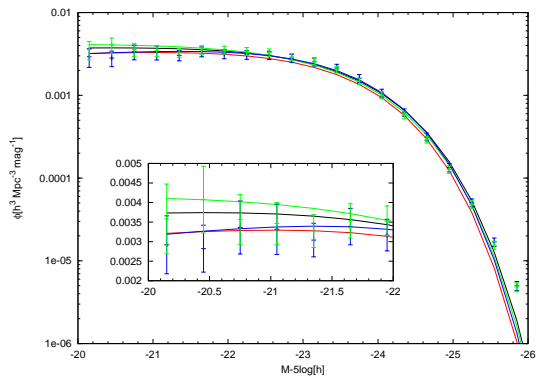}}}
\caption{The Schechter function for the four patches A (black), B (red), C (green) and D (blue), taking the largest magnitude cut $K_{\rm s} < 13.5{\rm mag}$. Although the Schechter functions exhibit some deviation from anisotropy, the deviation is not statistically significant. Also shown is the non-parametrically estimated luminosity function for patches C (green) and D (blue), with $1-\sigma$ error bars estimated from our $N_{\rm real} =1000$ realisations. }
\label{fig:a} 
\end{figure}

\begin{figure*}
\centering
\mbox{\resizebox{0.45\textwidth}{!}{\includegraphics[angle=0]{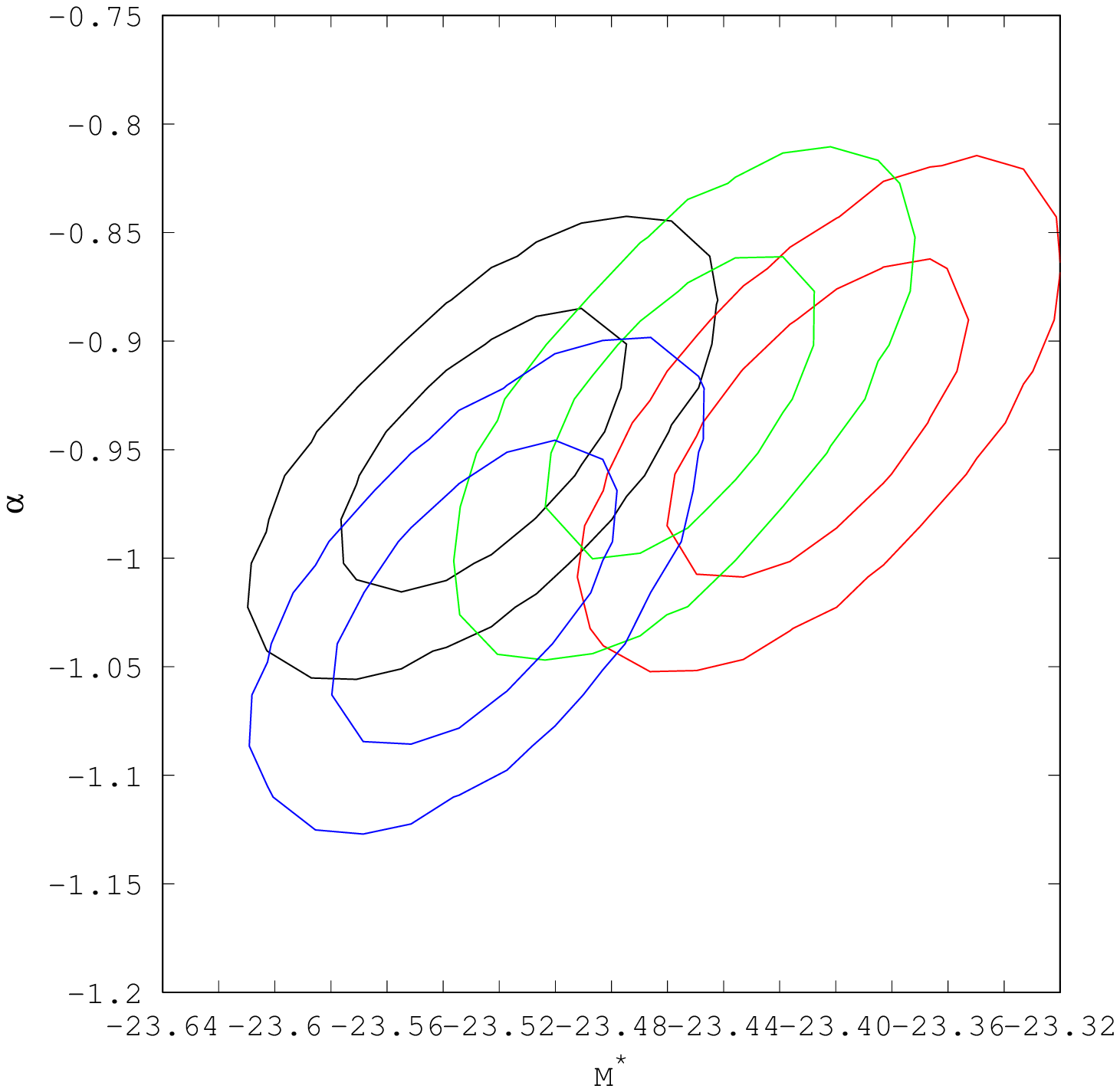}}}
%\mbox{\resizebox{0.45\textwidth}{!}{\includegraphics[trim=0mm 0mm 60mm 150mm,clip]{test_2D.eps}}}
\mbox{\resizebox{0.45\textwidth}{!}{\includegraphics[angle=0]{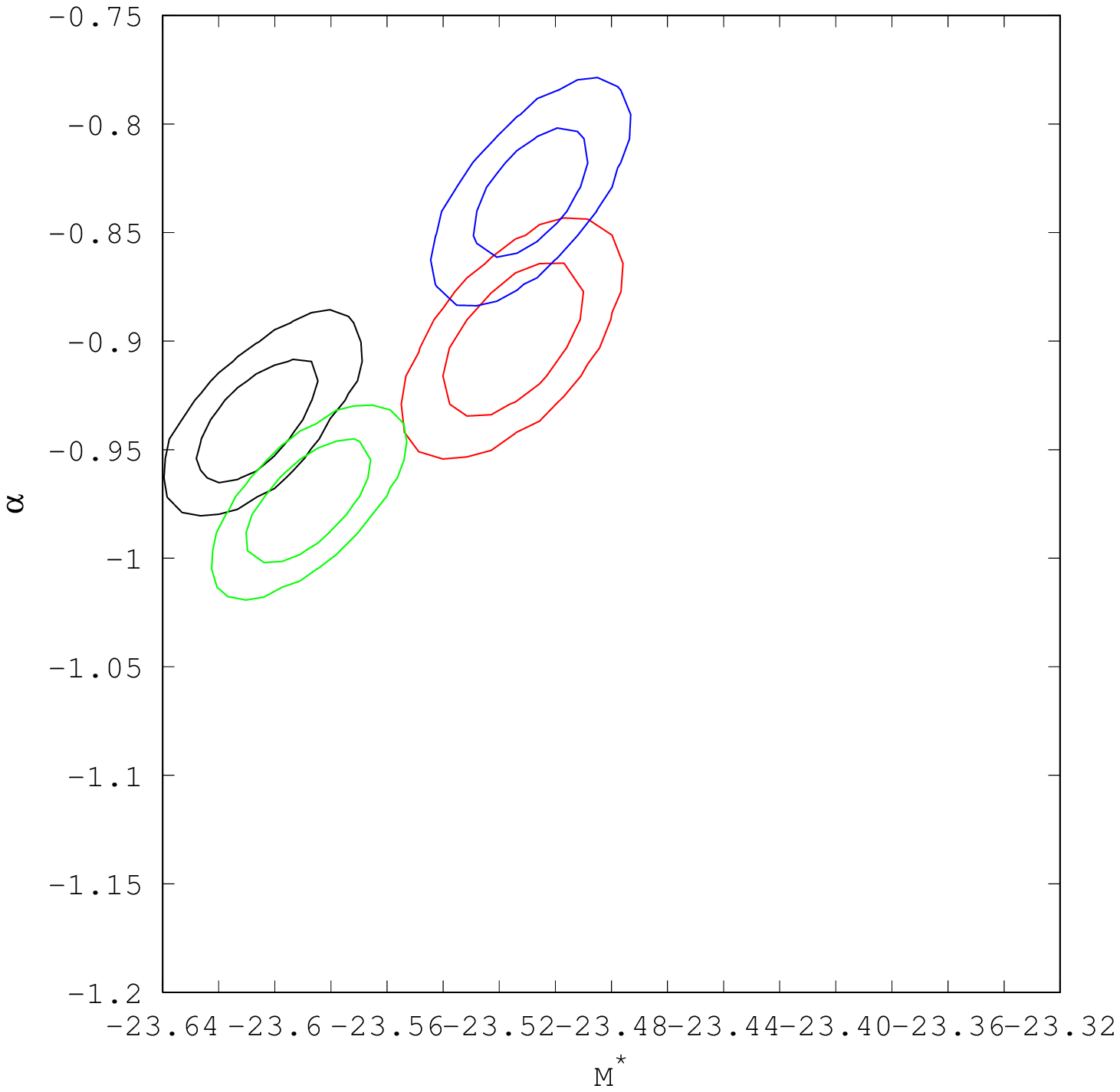}}}
\caption{The estimated $1,2-\sigma$ errors on the luminosity function parameters $(\alpha,M^{*})$ obtained using the STY maximum likelihood method. We consider two magnitude cuts - $K_{\rm s} < 12.5$ [Left panel] and $K_{\rm s} < 13.5$ [Right panel]. The different coloured contours correspond to the four patches A (black), B (red), C (green) and D (blue). The error bars are consistently higher for the lower magnitude cut, consistent with the fact that we are probing a smaller volume in the left panel. In the right panel we observe a dichotomy between the north and south galactic planes, however as argued in the text this apparent discrepancy is likely due to the relatively poor fit of the parametric form to the data.
}
\label{fig:4} 
\end{figure*}

We begin by presenting the best fit values for $(\alpha,M^{*})$ for each patch and for each magnitude cut in table \ref{tab:1}, and exhibit the luminosity functions graphically in fig.\ref{fig:a}. The values of $M^{\ast}$ and $\alpha$ exhibit some asymmetry with respect to position on the sky. At high magnitude cut $K_{\rm s} < 13.5$, the `knee' parameter $M^{\ast}$ takes marginally brighter values in the north galactic plane, and also shows signs of evolution with magnitude cut (corresponding to increasing the redshift of the sample). In fig.\ref{fig:4} we exhibit the $(\alpha,M^{\ast})$ contours for the four patches, which again suggests some form of anisotropy in the Schechter function. The quoted errors in table \ref{tab:1} - $\delta \alpha$, $\delta M^{\ast}$ - are the one dimensional marginalized errors on $\alpha,M^{\ast}$. One might be tempted to infer from the best fit values of $M^{\ast}$ and its associated error $\delta M^{\ast}$ that there is a statistically significant dipole in the north/south galactic plane. However, there are two possible sources of error that might lead to a spurious result - one is the large uncertainties associated with the photometric redshift errors which must be accounted for. The second is the quality of fit of the Schechter function in each patch, which must be assessed.

\begin{center}
\begin{table}
\begin{tabular}{ccc}
\toprule 
 $K_{\rm s} ({\rm mag})$   & $\alpha$ & $M^{*}-5\log[h]$  \\
    \midrule
$12.50$     & -0.96    &-23.48      \\
$13.00$     & -0.92    &-23.49    \\
$13.50$     & -0.92    &-23.56   \\
    \bottomrule
\end{tabular}
\caption{The best fit values of the Schechter function parameters $(\alpha,M^{*})$, obtained from the whole sky. We make three successive magnitude cuts $K_{\rm s} < (12.5,13.0,13.5) {\rm mag}$. As for the patches, we find an increasing value of $|M^{*}|$ with magnitude cut (corresponding to larger volumes being surveyed).}
\label{tab:4}
\end{table}
\end{center}

\begin{figure}
\centering
\mbox{\resizebox{0.4\textwidth}{!}{\includegraphics[angle=270]{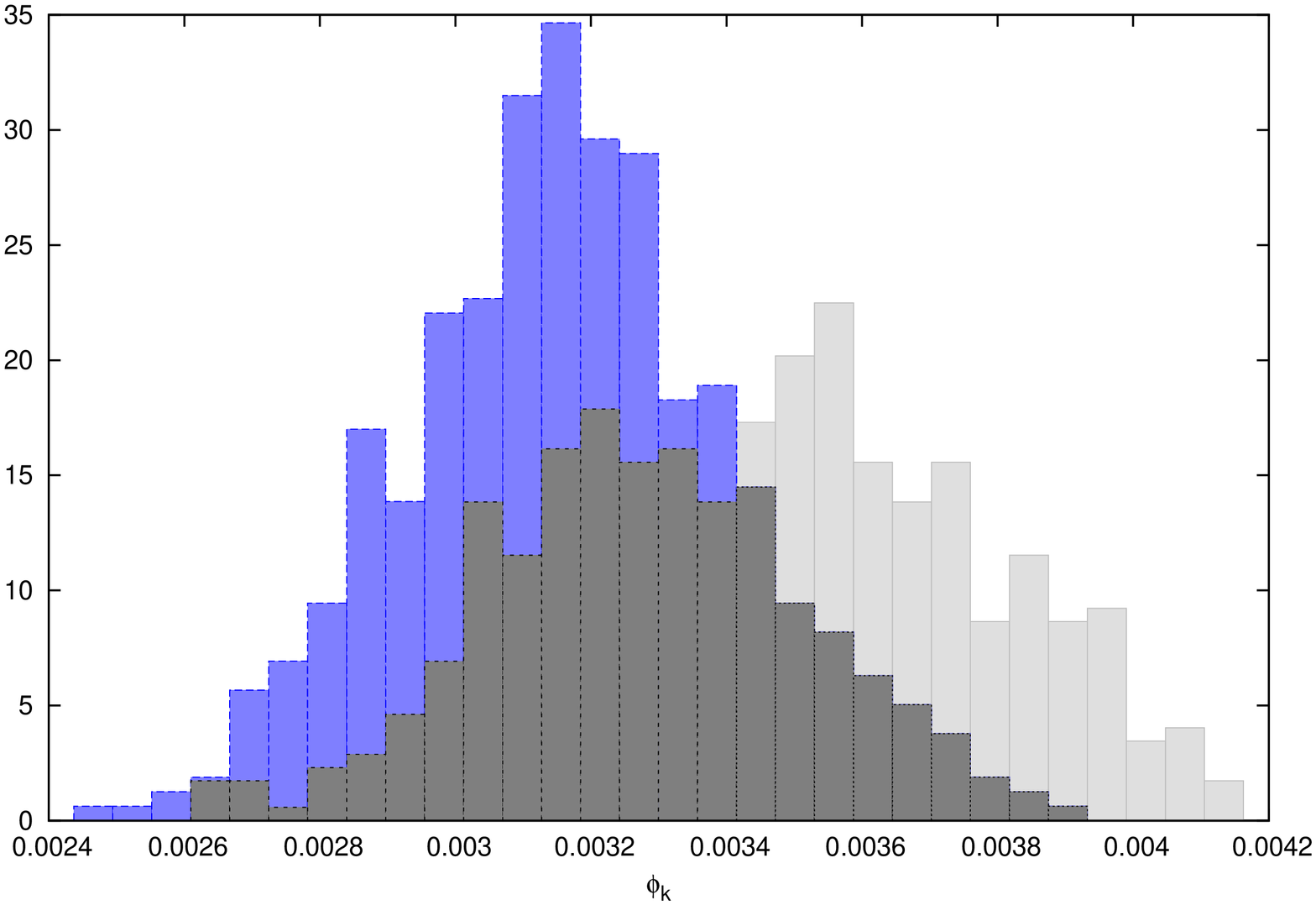}}}
\caption{An example of our $p_{\rm ij}$ measure. From the $N_{\rm real}=1000$ realisations of the $k^{\rm th}$ magnitude bin, we extract empirical probability distribution functions of the luminosity function $\phi_{\rm k}$. The $p_{\rm ij}$ value between the $i^{\rm th}$ and $j^{\rm th}$ patches is given by half of the combined area under both PDF's. Here we show the distribution of $\phi_{\rm k}$ for patches C and D in blue and light grey respectively. The dark grey region is the shared area under both of the distributions - half of the dark grey area corresponds to $p_{CD}$ for this magnitude bin.  
}
\label{fig:gau} 
\end{figure}

To address both of these problems, we estimate the luminosity function non-parametrically using the EEP method outlined in section \ref{sec:2}. We create $N_{\rm bin} = 20$ magnitude bins equally spaced in the range $M=(-20,-26){\rm mag}$, and construct $\phi_{\rm k}$ (with $k=(1,N_{\rm bin})$) for the $N_{\rm real} = 1000$ mock data sets. In each magnitude bin, we thus obtain a probability distribution for $\phi_{\rm k}$ due to the uncertainty in the underlying redshift of the galaxy sample. One can quantify the statistical significance of any anisotropic signal by comparing the $\phi_{\rm k}$ distributions in different patches. If we denote the magnitude bin with label ${\rm k}$ (which runs from ${\rm k}=1,N_{\rm bin}$), then our measure of anisotropy between patch $i$ and patch $j$ (where $i,j$ run over A,B,C,D) is $p_{\rm ij}^{\rm k}$, which is half of the combined area under the probability distributions for $\phi_{\rm k}^{(i)}$ and $\phi_{\rm k}^{(j)}$. A schematic of $p_{\rm ij}^{\rm k}$ for one particular magnitude bin is displayed in fig.\ref{fig:gau}. The two distributions correspond to the $\phi_{\rm k}$ obtained in a common magnitude bin, constructed from the $N_{\rm real} =1000$ realisations. The two distributions here are from patches A and D, and $p_{\rm ij}^{\rm k}$ is equivalent to half of the dark grey shaded area. As we do not know the `true' redshifts of our galaxy sample, in each patch and each magnitude bin we are actually constructing a distribution of possible Luminosity Functions from our $N_{\rm real}=1000$ realisations. One can therefore interpret the $p_{\rm ij}$ quantity as a p-value, indicating the separation of the mean values of distributions in different patches. As such, we adopt the standard convention with p-values and take small values $p_{\rm ij}^{\rm k}\sim 10^{-2}$ to indicate a significant anisotropic signal in the data. 

We adopt two measures in this work - one is the average $p_{\rm ij}$ per magnitude bin,

\begin{equation} \bar{p}_{\rm ij} = {\sum_{k=1}^{N_{\rm bin}} p_{\rm ij}^{k} \over N_{\rm bin} } \end{equation} 

\noindent where we sum over $N_{\rm bin} = 20$ bins in the magnitude range $M = (-20,-26){\rm mag}$. Our second measure is simply the minimum value of $p_{\rm ij}^{\rm k,min}$ of the $N_{\rm bin}=20$ magnitude bins. We denote these quantities as $\bar{p}_{\rm ij}$ and $p^{\rm s}_{\rm ij}$ respectively. 

The degree of overlap of the $\phi_{\rm k}$ distributions in each bin will depend on the overall normalisation of the luminosity function, which in turn depends on the galaxy density in a given patch. We are attempting to test the isotropy of the distribution and disentangle the effects of anisotropy and inhomogeneity. For this purpose we normalize the luminosity function to unity over the absolute magnitude range considered here - $M=(-20,-29){\rm mag}$. We do this at each magnitude cut and in all patches. 

In tables \ref{tab:2b},\ref{tab:2} we exhibit the $\bar{p}_{\rm ij}$ and $p^{\rm s}_{\rm ij}$ values for A,B,C,D. The values in each parenthesis correspond to magnitude cuts $K_{\rm s} < (12.5,13.0,13.5) {\rm mag}$. The average $\bar{p}_{\rm ij}$ exhibit no statistically significant anisotropy between any patch, for any magnitude cut. The quantity $p^{\rm s}_{\rm ij}$ is smaller, being of order $\sim {\cal O}(0.1)$ for certain magnitude cuts and regions. However there is no strong evidence of anisotropy, which would be characterized by $p^{\rm s}_{\rm ij} \sim 10^{-2}$ values. 

The width of the $\phi_{\rm k}$ distribution in each magnitude bin is determined by the uncertainties in the photometric redshifts - larger errors on the measurements will have the effect of widening the $\phi_{\rm k}$ distributions and therefore increasing the $p_{\rm ij}$ values. On the evidence of tables \ref{tab:2b} and \ref{tab:2} we can conclude that given the uncertainties on our galaxy redshift measurements, the shape of the luminosity function is consistent with the assumption of isotropy over the magnitude range and sky area considered here. 

Our result is reflected in fig.\ref{fig:a}, where we exhibit the best fit Schechter functions for patches A-D for magnitude cut $K_{\rm s} < 13.5 {\rm mag}$. The Schechter function appears to show some evidence of a dichotomy in the north/south galactic plane in the faint end slope of $\phi(M)$. However in fig.\ref{fig:a} we also exhibit the non-parametric luminosity function reconstruction for patches C and D (shown as green and blue data points respectively). The associated errors are the $1-\sigma$ uncertainties obtained from the mock realisations. One can conclude that the luminosity function is consistent with isotropy, and the apparent Schechter function anisotropy is not significant.

Given that we have two estimates of the luminosity function, we can estimate the goodness of fit of the Schechter function in each patch by constructing a simple $\chi^{2}$ statistic using the non-parametrically estimated luminosity function obtained with the EEP approach. We take 

\begin{equation} \chi^{2} = \sum_{i=1}^{N_{\rm bin}} {\left(\phi_{\rm Schech, i} - \phi_{\rm EEP, i}\right)^{2} \over \sigma_{2, i}^{2}} \end{equation}

\noindent where $\phi_{\rm Schech, i}$ is the Schechter function ($\ref{eq:4}$) evaluated at the ${\rm i}^{\rm th}$ magnitude bin center, $\phi_{\rm EEP, i}$ is the non-parametrically estimated luminosity function, and $\sigma_{2, i}$ is the $95\%$ confidence limit obtained from the $N_{\rm real} = 1000$ mock data realisations in the $i^{\rm th}$ bin. We perform this simple fit for all $N_{\rm bin}=20$ bins in the range $M = (-20,-26){\rm mag}$ - the $\chi^{2}$ for each patch and each magnitude cut are presented in table \ref{tab:1}. It is clear that the ability of the Schechter function to fit the data varies significantly between the four patches. The fit also degrades considerably with increasing magnitude cut, indicating a failure to simultaneously fit the faint end flattening and bright end decay out to $K_{\rm s} < 13.5{\rm mag}$. Variations in this simple $\chi^{2}$ statistic vary by $\sim 40\%$ from patch to patch, suggesting a strong variability in the fit quality of the Schechter function in different regions of the sky.

\begin{center}
\begin{table}
\begin{tabular}{ccccc}
\toprule 
Patch & A & B & C &  D \\
A     &    & (0.28,0.31,0.31)    &   (0.30,0.29,0.21)  & (0.29,0.30,0.34)    \\
B     &    &   &   (0.34,0.35,0.33)  & (0.35,0.31,0.33)    \\
C     &    &     &    & (0.37,0.35,0.26)  \\
D     &     &     &     &     \\
\bottomrule
\end{tabular}
\caption{The average $p$-value per magnitude bin, $\bar{p}_{\rm ij}$, used as a statistical measure of anisotropy. The values in parenthesis correspond to magnitude cuts $K_{\rm s} < (12.5,13.0,13.5) {\rm mag}$. We find no significant deviation from isotropy between any of the four patches.}
\label{tab:2b}
\end{table}
\end{center}

\begin{center}
\begin{table}
\begin{tabular}{ccccc}
\toprule 
Patch & A & B & C &  D \\
A     &    & (0.12,0.13,0.18)    &   (0.22,0.19,0.10)  & (0.13,0.18,0.23)     \\
B     &     &    &   (0.15,0.24,0.09)  & (0.22,0.03,0.21)    \\
C     &    &    &     & (0.22,0.23,0.12)   \\
D     &     &   &     &     \\
\bottomrule
\end{tabular}
\caption{The smallest value of $\bar{p}^{\rm s}_{\rm ij}$ in any magnitude bin. $i,j$ denote the patch identifier $A-D$. As for $\bar{p}_{\rm ij}$, we again find little evidence of anisotropy. The successive values in each element of the table correspond to magnitude cuts $K_{\rm s} < (12.5,13.0,13.5) {\rm mag}$.}
\label{tab:2}
\end{table}
\end{center}

For large magnitude cut $K_{\rm s} < 13.5{\rm mag}$ we have a sufficiently large number of galaxies that we can further decompose the sky. We therefore create $N_{\rm patch} = 192$ equal area patches according to the HEALPIX, ${\rm NSide} = 4$ scheme. Each patch has area $A \sim 215 {\rm deg}^{2}$ and contains order $\sim {\cal O}(2000)$ galaxies with $K_{\rm s} < 13.5{\rm mag}$. In each patch we create $N_{\rm real} =1000$ realisations using the same procedure adopted previously. Once again, in each magnitude bin we again calculate the probability distribution for $\phi_{\rm k}$ and calculate the average $\bar{p}_{\rm ij}$ per bin. Now the indices ${\rm i,j}$ run over the $N_{\rm pix} = 112$ pixels that lie above our magnitude cut $|b| < 20^{\circ}$.

\begin{figure}
\centering
\mbox{\resizebox{0.4\textwidth}{!}{\includegraphics[angle=270]{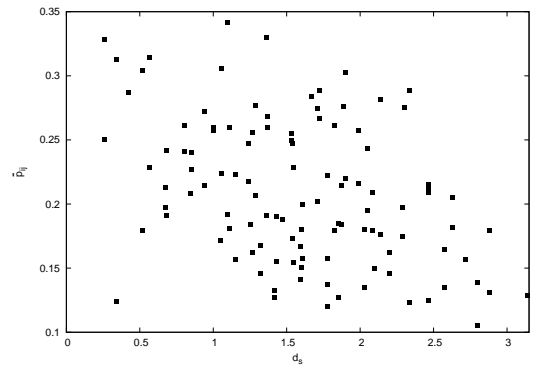}}}
\caption{For each of the $j =1,112$ patches with $|b|>20$, we calculate the $\bar{p}_{\rm ij}$ value between it and the other $i=1,112$, $i \ne j$ patches. We then calculate the correlation coefficient between the $\bar{p}_{\rm ij}$ values and and distance between the $(i,j)$ pixels on the unit sphere. Exhibited here are the $\bar{p}_{\rm ij}$ values as a function of distance $d_{\rm s,ij}$ for the pixel $j$ with the largest correlation coefficient. The pixel center corresponds to $(b,\ell)=(-30^{\circ},135^{\circ})$. We see a clear trend of decreasing $p$-value with increasing distance between pixels. Such behaviour is indicative of a dipole in the galaxy distribution. The correlation coefficient is calculated as $r=-0.41$. 
}
\label{fig:the} 
\end{figure}

For each patch $i$, we calculate the $\bar{p}_{\rm ij}$ value for the remaining $j=112$ pixels with $j \ne i$. We search for correlations between $\bar{p}_{\rm ij}$ and the distance on the unit sphere between patches $i$ and $j$, which we denote $d_{\rm s, ij}$. A signal of statistical anisotropy - specifically a dipole - would manifest itself as a negative correlation between $\bar{p}_{\rm ij}$ and $d_{\rm s, ij}$. We quantify the degree of correlation between the two variables via the correlation coefficient 

\begin{equation} r_{\rm j} = {\sum_{i=1}^{N_{\rm pix}} (\bar{p}_{\rm ij} - \tilde{p}_{\rm ij})(d_{\rm s,ij} - \bar{d}_{\rm s,ij}) \over \sqrt{\sum_{i=1}^{N_{\rm pix}} (\bar{p}_{\rm ij} - \tilde{p}_{\rm ij})^{2}} \sqrt{\sum_{i=1}^{N_{\rm pix}} (d_{\rm s,ij} - \bar{d}_{\rm s,ij})^{2}}} \end{equation}

\noindent where $\tilde{p}_{\rm ij}$ and $\bar{d}_{\rm s,ij}$ are the average values of these quantities calculated for all $i = (1,112)$ pixels with respect to pixel $j$ (with $i \ne j$). In fig.\ref{fig:the} we exhibit the $\bar{p}_{\rm ij}$ values as a function of $d_{\rm s}$ for the pixel with the maximum $|r_{\rm j}|$ value. The pixel $j$ has $r_{\rm j} = -0.41$ and is located at $(b,l) = (-30^{\circ},135^{\circ})$ in galactic coordinates. One can clearly see that while none of the $\bar{p}_{\rm ij}$ values individually exhibit strong statistical evidence for anisotropy - $\bar{p}_{\rm ij} > 0.1$ in all cases - there is a consistent trend of smaller $\bar{p}_{\rm ij}$ values for larger distances between the patches. This indicates weak evidence for a dipole in the shape of the luminosity function, in the direction $(b,\ell) = (-30,135)$ and $(b,\ell) = (30,315)$. The direction of this dipole is qualitatively consistent with that found in previous works \cite{Gibelyou:2012ri}.

\begin{figure}
\centering
\mbox{\resizebox{0.4\textwidth}{!}{\includegraphics[angle=270]{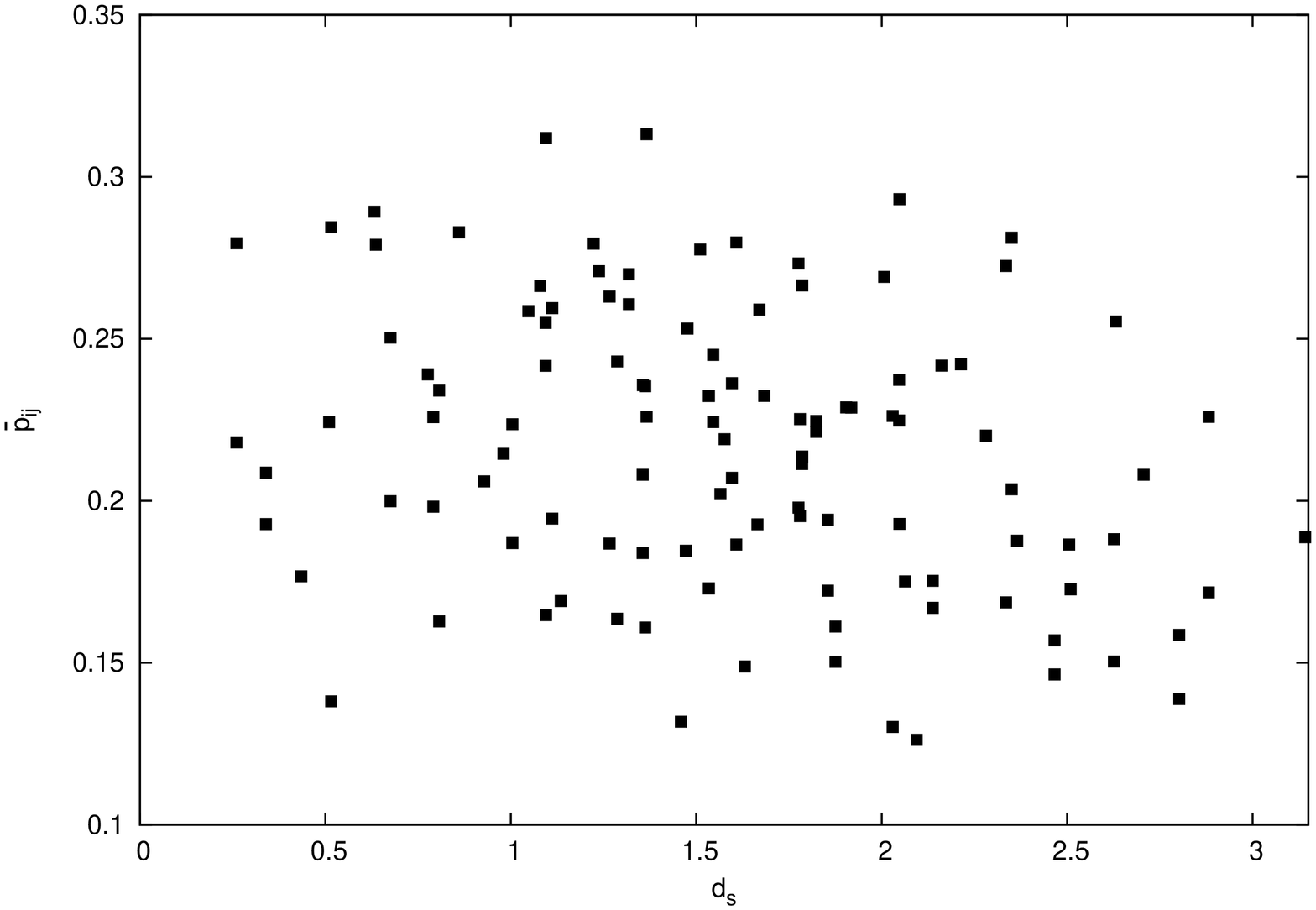}}}
\caption{Exhibited are the $\bar{p}_{\rm ij}$ values as a function of distance $d_{\rm s}$ for the pixel with center closest to the equatorial north pole. We see no indication of correlation between $\bar{p}_{\rm ij}$ and $d_{\rm s}$, indicating that there is no significant bias in the catalog due to photometric calibration differences in the north/south sky.
}
\label{fig:equ} 
\end{figure}

As an additional check that no anisotropic biases appear in the galaxy distribution due to different photometric calibrations in the equatorial plane, we also calculate the $\bar{p}_{\rm ij}$ and $r_{\rm j}$ values for the patch on the sky whose center lies closest to the equatorial north pole. Labeling this pixel as $j_{\rm eq}$, we exhibit $\bar{p}_{\rm ij_{\rm eq}}$ between this patch and the remaining $N=111$ pixels in fig.\ref{fig:equ}. The correlation coefficient for this pixel is lower, $r_{\rm j_{\rm eq}}=-0.24$, which indicates only very weak evidence for an anti-correlation between $\bar{p}_{\rm ij_{\rm eq}}$ and $d_{\rm s,ij_{\rm eq}}$. None of the $\bar{p}_{\rm ij_{\rm eq}}$ values are individually significant. We can conclude that the different photometric calibration in the north/south equatorial regions do not give rise to any statistically significant anisotropy.

\section{\label{sec:5}Discussion and Conclusion} 

In this work we have studied the properties of the local Universe using low redshift galaxy data. In doing so, we have found that there is some evidence that the shape of the luminosity function is asymmetric on the sky, with a dichotomy in the north/south galactic plane. When constructing the luminosity function using a parametric estimator, the parameters dictating the shape of the luminosity function exhibit systematically lower values in the north galactic plane. 

However, by utilising a non-parametric estimator to reconstruct the binned luminosity function, we have found that there is no statistically significant evidence for anisotropy when we account for photometric redshift uncertainty. The individual $p$-values obtained between any two patches on the sky indicate no evidence for a directional dependence of the shape of the luminosity function. However, we do observe some suggestion of a dipole in the north/south galactic plane, with a maximal signal at galactic coordinates $(b,l)=(30^{\circ},315^{\circ})$ (or equivalently $(b,l)=(-30^{\circ},135^{\circ})$). This result is consistent with existing works \cite{Gibelyou:2012ri}, which use a different method to estimate the anisotropy in the local number counts. 

Our results are furthermore in agreement with \cite{Feix:2014bma}, in which the local bulk velocity $V_{\rm bulk}$ was probed by calculating the effect of $V_{\rm bulk}$ on the galaxy luminosity function. The authors were able to reconstruct the magnitude and direction of $V_{\rm bulk}$ using galaxies from the SDSS data release 7 \cite{Feix:2014bma}, finding $|V_{\rm bulk}| = 120 \pm 115$ and $355 \pm 80 {\rm km s^{-1}}$ in redshift bins $0.02 < z < 0.07$ and $0.07 < z < 0.22$, with direction $(l,b) \simeq (310^{\circ},-25^{\circ}) \pm (30^{\circ},10^{\circ})$ and $(310^{\circ},5^{\circ}) \pm (10^{\circ},15^{\circ})$ respectively. Similarly, we only find an anisotropic signal when using our largest magnitude cut $K_{\rm s} < 13.5 {\rm mag}$ (corresponding to a higher redshift sample $z \lesssim 0.25$) and a direction that is consistent within the errors quoted. Since we do not attempt to model the effect of $V_{\rm bulk}$ on the Luminosity Function explicitly, we can make no claim as to the magnitude of $|V_{\rm bulk}|$ using our method, which should be interpreted as a null test of the isotropic hypothesis. 

We expect that using future spectroscopic data will allow us to increase the statistical significance of our finding. The lack of `true' redshifts for a large fraction of our galaxy sample forced us to construct a distribution of possible Luminosity Functions, based on realisations in which we inferred the spectroscopic redshifts from probability distribution functions of $z_{\rm spec} - z_{\rm photo})$ obtained from the data. The photometric redshift errors are the overwhelmingly dominant source of uncertainty in our calculation and increasing the number of spectroscopic redshifts available will have the effect of narrowing the Luminosity Function distributions, leading to increased sensitivity of $p_{\rm ij}$ to anisotropy. We note that related work has studied the effect of photometric redshift uncertainty on estimations of the Luminosity Function - see for example \cite{Ramos:2011nd}. 

If our result is taken in conjuncture with other recent work \cite{Keenan:2012gr,Keenan:2013mfa}, it is clear that there is some evidence that the local distribution of galaxies is neither homogeneous or isotropic. However, the extent to which this conclusion will impact cosmological observables is not clear. A detailed study of this question will remain for future work. It is also of considerable interest to test the consistency of the luminosity function in different regions of the sky using data sets that can probe a larger cosmological volume. Certainly, testing the scale at which the luminosity density convergences to its homogeneous asymptote remains a subject of considerable interest.

\acknowledgements{The authors would like to thank Eric Linder and Changbom Park for useful discussions, and for the detailed comments of Ryan Keenan and Maciej Bilicki that allowed us to significantly improve the paper. This publication makes use of data products from the Two Micron All Sky Survey, which is a joint project of the University of Massachusetts and the Infrared Processing and Analysis Center/California Institute of Technology, funded by the National Aeronautics and Space Administration and the National Science Foundation. Maps and results have been derived using the
HEALPix (http://healpix.jpl.nasa.gov) software package
developed by \cite{Gorski:2004by}. We thank the Wide Field Astronomy Unit at the Institute
for Astronomy, Edinburgh for archiving the 2MPZ catalog,
which can be accessed at http://surveys.roe.ac.uk/ssa/TWOMPZ. S.A.A and A.S wish to acknowledge support from the Korea Ministry of Education, Science and Technology, Gyeongsangbuk-Do and Pohang City for Independent Junior Research Groups at the Asia Pacific Center for Theoretical Physics. A.S. would like to acknowledge the support of the National Research Foundation of Korea (NRF-2013R1A1A2013795).}

\end{document}